\documentclass[onecolumn]{aastex63}

\newcommand{\sy}{\color{black}}
\usepackage{hyperref}
\usepackage{graphicx,times}
\usepackage{subfigure}
\newcommand{\be}{\begin{equation}}
\usepackage{threeparttable}
\usepackage{booktabs}
\newcommand{\ee}{\end{equation}}
\newcommand{\bea}{\begin{eqnarray}}
\newcommand{\eea}{\end{eqnarray}}

\usepackage{amsmath}
\usepackage{cases}
\usepackage{longtable}
\usepackage{hyperref}
\usepackage{epstopdf}
\usepackage{amsmath,bm}
\usepackage{amssymb}
\usepackage{natbib}
\usepackage{morefloats}
\usepackage{multirow}
\usepackage{array}
\usepackage{verbatim}
\usepackage{color}
\usepackage{lineno}

\begin{document}
\title{Damping of Alfv\'{e}n waves in MHD turbulence and implications for cosmic ray streaming instability and galactic winds}

\author{Alex Lazarian}
\affiliation{Department of Astronomy, University of Wisconsin, 475 North Charter Street, Madison, WI 53706, USA; 
lazarian@astro.wisc.edu}

\affiliation{Centro de Investigación en Astronomía, Universidad Bernardo O’Higgins, Santiago, General Gana 1760, 8370993,Chile}

\author[0000-0002-0458-7828]{Siyao Xu}
\affiliation{Institute for Advanced Study, 1 Einstein Drive, Princeton, NJ 08540, USA; sxu@ias.edu
\footnote{Hubble Fellow}}






\begin{abstract}
Alfv\'{e}nic component of MHD turbulence damps Alfv\'{e}nic waves. The consequences of this effect are important for many processes, from cosmic ray (CR) propagation to launching outflows and winds in galaxies and other
magnetized systems.
We discuss the differences in the damping of the streaming instability by turbulence and the damping of 
{\sy a plane parallel wave. 
The former takes place in the system of reference aligned with the local direction of magnetic field along which CRs stream. 
The latter is in the reference frame of the mean magnetic field and 
traditionally considered in plasma studies.
We also compare the turbulent damping of streaming instability with ion-neutral collisional damping, which becomes the dominant damping effect at a sufficiently low ionization fraction.
Numerical testing and astrophysical implications are also discussed. }
\end{abstract}

\section{Propagation of Alfv\'{e}n waves in MHD turbulence}

Astrophysical media are turbulent and magnetized (see a collection of relevant reviews in Lazarian et al. 2015a). The propagation of Alfv\'{e}n waves in turbulent magnetized media is an important astrophysical problem that influences fundamental astrophysical processes (see e.g., Uhlig et al. 2012, Wiener, Oh \& Guo 2013, van der Holst et al. 2014, Lynch et al. 2014).  This {\sy review} focuses on the damping of Alfv\'{e}n waves in MHD turbulence. The Alfv\'{e}n waves can arise from instabilities induced by cosmic rays (CRs), e.g. from the streaming of CRs (Lerche 1967, Kulsrud \& Pearce 1969, Wentzel 1969, Skilling 1971), and the gyroresonance instability related to the compression of magnetic field and CRs (see Lazarian \& Beresnyak 2006).
{\sy They can also be} generated by large scale perturbations of magnetic field (see Konigl 2009 and ref. therein, Suzuki 2013). 

Turbulent damping of Alfv\'{e}n waves causes heating of, e.g. coronal gas in Solar atmosphere (e.g. Arber, Brady \& Shelyag 2016, Reep \& Russell 2016).
In the case of the streaming instability, turbulent damping 
{\sy suppresses its growth and affects the streaming speed of CRs.}
As a result, turbulent damping of streaming instability is important for studies on the diffusion and acceleration of CRs in shocks, galaxies, and galaxy clusters
(Bell 1978, Kulsrud 2005, Ensslin et al. 2011, Blasi et al. 2012, Wiener et al. 2013, Badruddin, \& Kumar, A. 2016, Xu \& Lazarian 2022),
stellar wind launching (e.g. Suzuki \& Inutsuka  2005, van Ballegooijen, \& Asgari-Targhi 2016),
and 
galaxy evolution (e.g., Hopkins et al. 2021).

 

It should be noted that the well-known study of Alfv\'{e}n wave damping by turbulent plasmas performed by Silimon \& Sudan (1989) employed an unrealistic model of isotropic MHD turbulence. Later, turbulent damping of Alfv\'{e}n waves was mentioned as a process for suppressing CR streaming instability in Yan \& Lazarian (2002, henceforth YL02). This process was quantified by Farmer \& Goldreich (2004, henceforth FG04), where the {Goldreich \& Sridhar (1995; henceforth, GS95)} model of Alfv\'{e}nic turbulence {\sy with scale-dependent anisotropy} was adopted. The limitation of the aforementioned study was that for the calculations it was assumed that turbulence is injected isotropically with the turbulent velocity $u_L$ {\sy exactly} equal to the Alfv\'{e}n velocity $V_A$, i.e. Alfv\'{e}n Mach number $M_A$ equal to unity. In addition, 
{\sy only turbulent damping of streaming instability was considered.}

 Following the study in Lazarian (2016), {\sy we will seperately discuss the turbulent damping of Alfv\'{e}n waves
 that are generated by streaming instability and by large-scale magnetic perturbations. 
 We will demonstrate the strong dependence of turbulent damping on $M_A$ in various turbulence regimes and astrophysical media with different levels of medium magnetization. }
 In \S 2 we provide the derivation of the Alfv\'{e}nic turbulent scaling. In \S 3 we describe the turbulent damping of Alfv\'{e}n waves generated by streaming instability in the reference system aligned with the local direction of turbulent magnetic field. {\sy In \S 4 we discuss the turbuelnt damping of Alfv\'{e}n waves
 induced by large-scale magnetic perturbations 
 in a global system of reference. 
 We compare the turbulent damping with ion-neutral collisional damping of streaming instability in a partially ionized medium in \S 5.} 
 The numerical testing of the theoretical predictions is provided in \S 6. The discussion of the astrophysical implications on propagation of CRs in galaxies and launching of winds follows in \S 7. The summary is given in \S 8. 

\section{Derivation of Alfv\'{e}nic turbulent scaling}

In Alfv\'{e}nic turbulence the relative perturbations of velocities and magnetic fields are related as
follows:
\begin{equation}
\frac{\delta B_l}{B}=\frac{\delta B_l}{B_L}\frac{B_L}{B}=\frac{u_l}{u_L} M_A=\frac{u_l}{V_A},
\label{same}
\end{equation}
where $B_l$ is the fluctuation of the magnetic field $B$ at scale $l$, $B_L$ is the fluctuation of the magnetic field at the driving scale $L$ of turbulence. 
Correspondingly, $u_l$ is the turbulent velocity fluctuation at the scale $l$ and $u_L$ is the turbulent velocity at $L$. {$M_A=u_L/V_A$ is the Alfv\'{e}n Mach number. }

One way to understand the non-linear interactions of Alfv\'{e}n waves within the MHD turbulent cascade is to consider colliding Alfv\'{e}n wave packets with parallel scales $l_{\|}$ and perpendicular scales $l_{\bot}$. The collision of a wave packet induces an energy change
\begin{equation}
 \Delta E \sim (du^2_l/dt) \Delta t,
 \label{init}
 \end{equation}
 where the term in brackets manifests the change of the energy of a wave packet induced by its interaction with the oppositely moving Alfv\'{e}n wave packet. {\sy The time of this interaction is equal to the time of the passage of these wave packets through each other}. As the size of the {\sy packet} is 
 $l_{\|}$, the interaction time is simply
 $\Delta t \sim l_{\|}/V_A$. 
 
 The rate of turbulent energy cascade is related to the rate of structure change of the oppositely moving wave packet. The 
 latter is $u_l/l_{\bot}$. As a result, Eq. (\ref{init}) provides
\begin{equation}
  \Delta E 
 \sim {\bf u}_l \cdot \dot{\bf u}_l\Delta t
 \sim  (u_l^3/l_{\perp}) (l_{\|}/V_A),
 \label{change}
\end{equation}

The fractional change of packet energy taking place per collision is $\Delta E/E$. This characterises the strength of the nonlinear turbulent interaction:
\begin{equation}
  f \equiv \frac{\Delta E}{u^2_l}
                           \sim \frac{ u_l l_{\|} }{ V_A l_{\perp} }.
                         \label{fraction}
\end{equation}
In Eq. (\ref{fraction}), $f$ is the ratio of the shearing rate of the wave packet, i.e.
$u_l/l_{\bot}$, to its propagation rate, i.e. $V_A/l_{\|}$.

One can identify two distinct cases. If $f\ll 1$, the shearing rate is significantly smaller than the propagation rate, and  
the cascade presents a random walk process. Therefore
\begin{equation}
\aleph=f^{-2}
\label{aleph}
\end{equation}
steps are required for the energy cascade, {\sy and therefore the cascading time is} 
\begin{equation}
t_{cas}\sim \aleph \Delta t .
\label{tcas}
\end{equation}
$\aleph>1$ corresponds to the {\it weak} turbulent cascade. Naturally, $\aleph$ cannot become less than unity. Therefore, the limiting case is $\aleph\approx 1$. This is the case of  {\it strong} MHD turbulence. 

 
{Traditionally,} the wavevectors are defined in the system of reference related to the mean field.
{However,}
 the system of reference related to a wave packet with given parallel and perpendicular dimensions is more
 relevant when dealing with strong MHD turbulence. We take this into account by considering Alfv\'{e}n wave packets having the 
 dispersion relation $\omega = V_A |k_{\|}|$, 
 where we use $k_{\|}\sim l_{\|}^{-1}$ as the component of wavevector
parallel to the local background magnetic field. As the result of interaction the increase of $k_{\bot}\sim l_{\bot}^{-1}$ occurs. In the
rest of the discussion we use $l_\|$ and $l_\bot$ that are defined in the local frame of wave packets.

{In weak turbulence,}
the decrease of 
$l_{\bot}$ while $l_{\|}$ does not change signifies the increase of the energy change per collision. This forces
$\aleph$ to be of the order of unity. In this case one gets
\begin{equation}
u_l l_{\bot}^{-1}\approx V_A l_{\|}^{-1}
\label{crit}
\end{equation}
{in strong turbulence,}
which signifies the cascading time being equal to the wave period $\sim \Delta t$. Any further decrease of $l_{\bot}$ inevitably results in
the corresponding decrease of $l_{\|}$ and Eq. (\ref{crit}) 
{is still} satisfied. The change of $l_{\|}$ entails the increase of  
the frequencies of interacting waves. This is compatible with the conservation of energy condition above, as the 
cascade introduces the uncertainty in
wave frequency $\omega$ of the order of $1/t_{cas}$. 

The cascade of turbulent energy satisfies the relation (Batchelor 1953):
\begin{equation}
\epsilon\approx u_l^2/t_{cas}=const,
\label{cascading}
\end{equation}
which for the hydrodynamic cascade provides
\begin{equation}
\epsilon_{hydro}\approx u_l^3/l\approx u_L^3/L=const,
\end{equation}
where the relation for the cascading time $t_{cas}\approx l/u_l$ is employed.

For the weak turbulent cascade with $\aleph \gg 1$, we have (LV99)
\begin{equation}
\epsilon_w \approx \frac{ u_l^4} {V_A^2 \Delta t (l_{\bot}/l_{\|})^2} \approx \frac{u_L^4}{V_A L},
\label{eps_weak}
\end{equation}
where Eqs. (\ref{cascading}) and (\ref{tcas}) are used. The isotropic turbulence injection at scale $L$ results in the second relation in Eq. (\ref{eps_weak}). 
Taking into account that for the weak turbulence
$l_{\|}$ is constant, it is easy to see that Eq. (\ref{eps_weak}) provides
\begin{equation}
u_l\sim u_L (l_{\bot}/L)^{1/2},
\label{u_weak}
\end{equation}
which is different from the hydrodynamic $\sim l^{1/3}$ scaling.\footnote{Using the relation $k E(k) \sim u_k^2$ it is easy to show that the energy spectrum of weak turbulence is $E_{k, weak}\sim k_{\bot}^{-2}$ (LV99, Galtier et al. 2000).} 

It was shown in LV99 that for turbulence with isotropic injection at scale $L$ with $V_L<V_A$ the transition to the strong regime corresponding to $\aleph\approx 1$ happens at the scale
\begin{equation}
l_{trans}\sim L(u_L/V_A)^2\equiv L M_A^2.
\label{trans}
\end{equation} 
As a result, the inertial range of weak turbulence is limited,  i.e. $[L, L M_A^2]$, and at $l_{trans}$ the turbulence transits into the regime of strong MHD turbulence. At the transition, the velocity is 
\begin{equation}
u_{trans}\approx V_A \frac{l_{trans}}{L}\approx V_A M_A^2,
\label{vtrans}
\end{equation}
which follows from $\aleph\approx 1$ condition given by Eqs. (\ref{fraction}) and (\ref{aleph}). 

The scaling relations for the strong turbulence with $V_L<V_A$ can be easily obtained. The turbulence is strong and
cascades over one wave period, which according to Eq. (\ref{crit}) is equal to $l_{\bot}/u_l$.  
Substituting the latter in Eq. (\ref{cascading}) one gets
\begin{equation}
 \epsilon_s\approx \frac{u_{trans}^3}{l_{trans}}\approx \frac{u_l^3}{l}=const. 
 \label{alt_sub2}
\end{equation}
The latter energy cascading rate is analogous to that in an ordinary hydrodynamic Kolmogorov cascade. However, this cascading takes place in
the direction perpendicular to the {\it local} direction of the magnetic field. \footnote{There is an intuitive way of presenting the Alfv\'{e}nic cascade in terms of eddies mixing the magnetic field in the direction perpendicular to the magnetic field surrounding the eddies. The existence of such magnetic eddies is possible due to the fact that, as shown in LV99, the turbulent magnetic reconnection happens within one eddy turnover. As a result, the existence of magnetic field does not constrain magnetic eddies, if they are aligned with the magnetic field in their vicinity, i.e. with the {\it local} magnetic field. This eddy representation of MHD turbulence vividly demonstrates the importance of the {\it local} system of reference, where $l_\bot$ and $l_\|$ are defined.}

This strong MHD turbulence cascade starts at $l_{trans}$ and
its injection velocity is given by Eq. (\ref{vtrans}). This provides another way to obtain the Alfv\'{e}nic turbulent scaling in strong turbulence regime (LV99)
\begin{equation}
u_{l}\approx V_A \left(\frac{l_{\bot}}{L}\right)^{1/3} M_A^{4/3},
\label{vll}
\end{equation}
which can be rewritten in terms of the injection velocity $u_L$ (see Eq. (\ref{vll}) )
\begin{equation}
\delta u_{l}\approx u_L \left(\frac{l_{\bot}}{L}\right)^{1/3} M_A^{1/3}.
\label{alternative}
\end{equation}
Substituting this in Eq. (\ref{crit}) we get the relation between the parallel and perpendicular
scales of the eddies (LV99):
\begin{equation}
l_{\|}\approx L \left(\frac{l_{\bot}}{L}\right)^{2/3} M_A^{-4/3}.
\label{Lambda1}
\end{equation}
The relations Eq. (\ref{Lambda1}) and (\ref{vll}) reduce to the GS95 scaling for transAlfv\'{e}nic turbulence if 
$M_A\equiv 1$. 

In the opposite case we deal {with {\it superAlfv\'{e}nic} turbulence, i.e. with $u_L > V_A$.} As a result, at scales close to the injection scale the turbulence is essentially hydrodynamic
as the influence of magnetic forces is marginal. Therefore, the velocity is Kolmogorov
\begin{equation}
u_l=u_L (l/L)^{1/3}.
\label{u_hydro}
\end{equation}
The magnetic field becomes more important at smaller scales and the  cascade changes its nature at the scale 
\begin{equation}
l_{A}=LM_A^{-3},~~~ M_A>1,
\end{equation}
at which the turbulent velocity becomes equal to the Alfv\'{e}n velocity (Lazarian 2006).  
The rate of cascade for $l<l_A$ is:
\begin{equation}
\epsilon_{superA}\approx u_l^3/l\approx M_A^3 V_A^3/L=const. 
\label{super_alt1}
\end{equation}
Unlike the case of subAlfv\'{e}nic turbulence, the case of superAlfv\'{e}nic turbulence can be reduced to the case of transAlfv\'{e}nic turbulence, but 
with $l_A$ acting as the injection scale. At scales $l<l_A$ 
\begin{equation}
l_{\|}\approx L \left(\frac{l_{\bot}}{L}\right)^{2/3} M_A^{-4/3},
\label{Lambda}
\end{equation}
\begin{equation}
u_{l}\approx u_{L} \left(\frac{l_{\bot}}{L}\right)^{1/3} M_A^{1/3}.
\label{vl}
\end{equation}
The relations for subAlfv\'{e}nic and superAlfv\'{e}nic tubulence that we obtain above coincide with the expressions first obtained in Lazarian \& Vishniac (1999) using {\sy a different approach}. These expressions will be used below in our discussion on turbulent damping of Alfv\'{e}n waves.

\section{Turbulent damping of streaming instability}

Linear Alfv\'{e}n waves undergo non-linear cascading when they propagate through Alfv\'{e}nic turbulence. This process is of MHD nature and the non-linear damping of Alfv\'{e}n waves does not depend on plasma microphysics.  
{The interaction between CR-driven Aflven waves and turbulence is similar to that of oppositely moving wave packets of turbulent cascade. }


The Alfv\'{e}n waves emitted parallel to the local magnetic field experiences the least distortions from the oppositely moving eddies. Thus the least distorted Alfv\'{e}n waves are those with the largest value of $l_{\bot}$. Indeed, the larger $l_{\bot}$, the longer time it takes for the evolution of the oppositely moving wave packets. For instance, for strong GS95 turbulence the {\sy time} corresponds to $l_{\bot}/v_l\sim l_{\bot}^{2/3}$. 

{The case of Alfv\'{e}n waves parallel to the local direction of magnetic field corresponds to streaming and gyroresonance instabilities.} In what follows, we will focus on the streaming instability.  
The dispersion of magnetic field directions with respect to the mean magnetic field determines the corresponding $l_\bot$. Naturally, the turbulent damping of Alfv\'{e}n waves is different for weak turbulence and strong turbulence.
{\sy Thus we will separately discuss turbulent damping of streaming instability in different turbulence regimes.}

\subsection{Streaming instability and local system of reference}
 
 {\sy The streaming instability of CRs happens as CR particles moving in one direction scatter back from a magnetic field perturbation and thus increase the amplitude of the perturbation.} 
 The induced perturbations are Alfv\'{e}n waves. If the Alfv\'{e}n waves are {\sy severely damped,} the CR particles can stream freely along the magnetic field. 
 
  Physically, the generation of Alfv\'{e}n waves takes place as CRs {\sy stream along} the local magnetic field. During the process the sampling scale for the magnetic field is the CR Larmor radius $r_L$. In this setting one should consider the process in the system of reference related to the local direction\footnote{The fact that MHD turbulence is formulated in terms of the local quantities is required for describing the interaction of MHD turbulence with CRs. Indeed, perturbations in the local system of reference are exactly what CRs interact with.} of the wondering magnetic field (LV99, Cho \& Vishniac 2000, Maron \& Goldreich 2001, Cho, Lazarian \& Vishniac 2002). 
 
 In the direction parallel to the local magnetic field, the growth rate of the streaming instability is (see Kulsrud \& Pearce 1969):
\begin{equation}
 \Gamma_{cr} \approx \Omega_B \frac{n_{cr}(>\gamma)}{n_i}\left(\frac{v_{stream}}{V_A} -1 \right),
 \label{gamm_cr}
 \end{equation}
 where $\Omega_B=eB/mc$ is the {\sy nonrelativistic} gyrofrequency, $n_{cr}$ is the number density of CRs with gyroradius $r_L>\lambda=\gamma mc^2/eB$, and $\gamma$ is the Lorentz factor.  
If the growth rate given by Eq. (\ref{gamm_cr}) is less than the rate of turbulent damping, {\sy the streaming instability is suppressed. }

\subsection{Damping by SubAlfv\'{e}nic strong turbulence}
Our first approach is based on calculating the distortion of Alfv\'{e}n waves by MHD turbulence as the waves propagate along magnetic field. The cause of the wave distortion is the {\sy field line wandering} over angle $\theta_x$. This angle is determined by the amplitude of magnetic field fluctuations $\delta B_x$ that {\sy are induced by }turbulent eddies with perpendicular scale $x$. One can see that the distortion induced during the time $t$ is
\begin{equation}
\delta_x\approx V_A t \sin^2\theta_x \approx V_A t \left(\frac{\delta B_x}{B} \right)_t^2, 
\label{deltax}
 \end{equation}
where the fluctuation induced by turbulence evolves as
\begin{equation}
\left(\frac{\delta B_x}{B}\right)_t\approx \left(\frac{u_x}{V_A}\right) \left(\frac{t}{x/u_x}\right).
\label{deltaB}
 \end{equation}
In the above expression $u_x$ denotes the velocity corresponding to the magnetic field fluctuation $\delta B_x$. The time $t$ in Eq. (\ref{deltaB}) is chosen to be less than the eddy turnover time $x/u_x$. As a result, the ratio reflects the partial sampling of the magnetic perturbation by the wave. By using the {\sy velocity scaling of strong subAlfv\'{e}nic turbulence} for $u_x$ in Eq. (\ref{deltaB}), it is easy to rewrite Eq. (\ref{deltax}) as
\begin{equation}
\delta_x\approx \frac{V_A^3 M_A^{16/3} t^3}{x^{2/3} L^{4/3}}.
\label{deltax2}
 \end{equation}

The wave damping corresponds to the ``resonance condition" $\delta_x=\lambda$, where $\lambda$ is the wavelength. Inserting this in Eq. (\ref{deltax2}) we obtain the
perpendicular scale of the ``resonance" magnetic fluctuations that distort the Alfv\'{e}n waves:
\begin{equation}
x\approx \frac{V_A^{9/2} t^{9/2} M_A^8}{\lambda^{3/2} L^2}.
\label{x1}
 \end{equation}
The time required to damp the Alfv\'{e}n waves is equal to the turnover time of the ``resonant" eddy:
\begin{equation}
t\approx\frac{x}{u_l}\approx\frac{V_A^2 t^3 M_A^4}{\lambda L}.
 \end{equation}
This provides the rate of non-linear damping of the Alfv\'{e}n waves,
\begin{equation}
\Gamma_{subA,s} \approx t^{-1},
\label{t}
 \end{equation}
or
\begin{equation}
\Gamma_{subA, s} \approx \frac{V_A M_A^2}{\lambda^{1/2} L^{1/2}},
\label{gamma1}
 \end{equation}
{\sy where the subsscript ``s" denotes ``strong turbulence".}
For transAlfv\'{e}nic turbulence, i.e. $M_A=1$, this result was obtained in FG04. The square of the Alfv\'{e}n Mach number dependence presented in Eq. (\ref{gamma1}) means a {\it significant} change of the damping rate compared to the transAlfv\'{e}nic case.\footnote{We note that in FG04 the injection scale for turbulence was defined not as the actual injection scale, but the scale at which the turbulent velocity becomes equal to the Alfv\'{e}n one. Such scale does not exist for subAlfv\'{e}nic turbulence.} 

If  the injection of turbulence is isotropic, the maximal perpendicular scale of strong subAlfv\'{e}nic motions is $x_{max}=LM_A^2$. Substituting this in Eq. (\ref{x1}) and using Eq. (\ref{t}) and Eq. (\ref{gamma1}) to express $t$, we get
\begin{equation}
\lambda_{max,s}\approx LM_A^4.
\label{max1}
 \end{equation}
The streaming CRs {\sy generate Alfv\'{e}n waves at a scale comparable to the  gyroradius $r_L$.} Thus it requires that
\begin{equation}
r_L<LM_A^4,
 \end{equation}
which is a notable limitation on CR energy if $M_A$ is small. 
{\sy The CRs with larger energies interact with weak turbulence as we will discuss in Section \ref{ssec:damsuba}. }

Due to the importance of turbulent damping of streaming instability, it is advantageous to provide another derivation of Eq. (\ref{gamma1}).  This alternative derivation is based on the picture of propagating wave packets that we used while obtaining Eq. (\ref{change}).  Consider two oppositely moving Alfv\'{e}n wave packets with the perpendicular scale $x'\sim k_{\bot}'^{-1}$. As we discussed earlier, each wave packet induces the distortion $\theta_x'$ of the oppositely moving waves. Consider an Alfv\'{e}n wave with wavenumber $k_{\|}^{-1}\sim \lambda$ moving parallel to the local direction of magnetic field. Such a wave is mostly distorted {\sy when interacting} with turbulent perturbations with 
the perpendicular wavenumber $k_{\bot} \sim k_{\|}\sin \theta_x'$. The interactions are most efficient if they are ``resonant", i.e. a wave with $k_{\bot}$ interacts with the oppositely moving packets and $k_{\bot}'=k_{\bot}$.\footnote{Simple estimates demonstrate that the interactions with smaller and larger turbulent scales are subdominant compared with the interaction with the ``resonant" scale.} Thus the perpendicular scale of the ``resonant" wave packet is  determined by the relation $k_{\|}\sin \theta_x=k_{\bot}$, which results in
\begin{equation}
\lambda \approx x \sin \theta_x\approx x \frac{\delta B_x}{B}.
\label{alt_lambda}
 \end{equation}
Using the scaling in Eqs. (\ref{vll}) and (\ref{deltaB}), we derive the ``resonant" perpendicular scale $x$:
\begin{equation}
x=L^{1/4} \lambda^{3/4} M_A^{-1}.
 \end{equation}
This can be used to determine the rate of damping defined as $\Gamma_{subA, s}\approx u_x/x$. This coincides with the earlier result given by Eq. (\ref{gamma1}). Then the maximal wavelength of the non-linearly damped Alfv\'{e}n waves can be obtained from Eq. (\ref{alt_lambda}) if the scale $l_{trans}$ is used instead of $x$, i.e.
\begin{equation}
\lambda_{max,s}\approx \left(\frac{u_{trans}}{V_A}\right) l_{trans}\approx LM_A^4.
\label{max_s}
 \end{equation}
Naturally, the latter coincides with the result given by Eq. (\ref{max1}). The minimal scale of non-linearly damped waves depends on
the perpendicular scale of the smallest Alfv\'{e}nic eddies $l_{min}$. The full range of $r_L$ for which turbulent damping is essential can be obtained by
 using Eq. (\ref{alt_lambda}) and the scaling of strong turbulence
given by Eq. (\ref{vll}):
\begin{equation}
\frac{l_{min}^{4/3}}{L^{1/3}}M_A^{4/3}<r_L<LM_A^4.
\label{min_max_s}
 \end{equation}
 The value of $l_{min}$ depends on the {\sy particular damping process of MHD turbulent cascade}, which can be relatively large in
a weakly ionized gas (see Xu \& Lazarian 2017). 
Due to the differences of $r_L$ for protons and electrons, Eq. (\ref{min_max_s}) presents
{\sy a possible situation when the streaming instability of CR electrons is not damped by turbulence, while it is damped for CR protons. }

We note that the turbulent damping of streaming instability for $r_L< \frac{l_{min}^{4/3}}{L^{1/3}}M_A^{4/3}$ is still present, although it is reduced. 
We can get an estimate of it by considering the distortion $\delta_x\ll \lambda$ given by Eq. (\ref{deltax2}) 
for the time period of
the wave $\lambda/V_A$.  This time is {\sy significantly less than the period of the eddy at the scale $l_{min}$, $t_{eddy}\approx 
l_{min}^{2/3} L^{1/3}/(V_A M_A^{4/3})$. The distortions act in a random walk fashion with the time step given by $t_{eddy}$.}
The damping requires $\lambda/\delta_x$ steps, which induces the damping rate
\begin{equation}
\Gamma_{sub, s, r_L\ll l_{min}}\approx \frac{M_A^{12} V_A r_L^4}{l_{min}^2 L^3}.
 \end{equation}
The latter clearly illustrates the inefficiency of damping when turbulence has the perpendicular scale larger than the ``resonant" scale.

\subsection{Damping by subAlfv\'{e}nic weak turbulence}\label{ssec:damsuba}

In many instances the weak turbulence is not important. It has a limited inertial range and transfers to strong turbulence at smaller scales. However, as we show below, this may not be true for wave damping by turbulence. For {\sy wavelengths longer} than $\lambda_{max, s}$ the wave is non-linearly damped through {\sy interactions with} the wave packets {\sy of the weak turbulence}, having perpendicular scales given by Eq. (\ref{alt_lambda}). Naturally, the scaling of weak turbulence given by Eq. (\ref{u_weak}) {\sy should} be used. This provides the relation between the Alfv\'{e}n wave wavelength and the perpendicular scale of the ``resonant" weak turbulence perturbation 
\begin{equation}
\lambda=l_\bot \left(\frac{l_{\bot}}{L}\right)^{1/2}M_A.
\label{lambda_w}
 \end{equation}
This delivers the perpendicular scale 
\begin{equation}
l_\bot\approx \lambda^{2/3} L^{1/3} M_A^{-2/3}.
\label{l_weak}
 \end{equation}

According to Eqs (\ref{aleph}) and (\ref{tcas}), the weak turbulence packets {\sy cascade} $\aleph$ times slower compared to the case of
strong turbulence. Taking into account that the parallel scale of weak turbulence wave packets is equal to the injection scale $L$, we have
\begin{equation}
\aleph \approx \left(\frac{V_A l_\bot}{u_l L}\right)^2.
 \end{equation}
 The rate of turbulent damping of the wave is therefore
\begin{equation}
\Gamma_{subA, w} \approx (\aleph \Delta t)^{-1}=\aleph^{-1}\frac{V_A}{L},
\label{gg}
 \end{equation}
which gives
\begin{equation}
\Gamma_{subA, w} \approx \frac{V_A M_A^{8/3}}{\lambda^{2/3} L^{1/3}},
\label{gamma_weak}
 \end{equation}
{\sy where the subsscript ``w" denotes ``weak turbulence".}
Note that compared to the case of damping  given by Eq. (\ref{gamma1}) {\sy we now have} a stronger dependence on $M_A$, 
as well as a different scaling with the wavelength $\lambda$.  

{\sy The maximal wavelength of the Alfv\'{e}n waves that is cascaded by the
weak cascade we derive by substituting $l_{\bot}=L$,} i.e. using the energy injection scale in Eq. (\ref{lambda_w}). This gives:
\begin{equation}
\lambda_{max, w}\approx L M_A.
\label{lambda_w_max}
 \end{equation}
{\sy Thus for CRs that generate Alfv\'{e}n waves, their $r_L$ should satisfy} 
\begin{equation}
LM_A^4<r_L<LM_A
\label{range_weak}
 \end{equation}
to interact with weak Alfv\'{e}nic turbulence. {\sy The underlying assumption here is that} $LM_A^4$ is larger than the turbulent damping scale $l_{min}$. Otherwise the lower boundary in Eq. (\ref{range_weak}) is determined by  $l_{min}$. 

Waves with $\lambda>\lambda_{max, w}$ can interact with the turbulent motions at the injection scale $L$. The cascade of such waves is induced
by the largest wave packets at a rate $\aleph^{-1} \frac{V_A}{L}$, i.e. 
\begin{equation}
\Gamma_{outer}\approx \aleph^{-1} \frac{V_A}{L} \approx M_A^2\frac{V_A}{L}, 
\label{outer}
 \end{equation}
 which does not depend on wavelength. Physically, this means that all waves in the range $LM_A<\lambda<L$ decay at the same rate that is determined by the restructuring of the magnetic field at the injection scale.
 
The above expression is valid for $\lambda<L$. In the case of $\lambda\gg L$ the rate {\sy is reduced due to the 
random walk}, which results in a factor $(L/\lambda)^2$, i.e.
\begin{equation}
\Gamma_{outer, extreme}\approx \aleph^{-1} \frac{V_A}{L} \frac{L^2}{\lambda^2}\approx M_A^2\frac{V_A}{L}\frac{L^2}{\lambda^2}.
\label{outer_extreme}
 \end{equation}
The latter result is relevant for the damping induced by turbulence injected at scales smaller than the wavelength.  

In terms of the dependence of damping rate on $\lambda$ for subAlfv\'{e}nic turbulence, we observe that the dependence becomes {\sy stronger} with the increase of $\lambda$ up to $\lambda=LM_A$. For $\lambda$ less than $LM_A^4$, {\sy the waves} interact with strong Alfv\'{e}nic turbulence and the damping rate $\Gamma$ is proportional to $\lambda^{-1/2}$. The scaling changes for waves longer than $LM_A^4$ but shorter than $LM_A$. The scaling of $\Gamma$ gets to $\lambda^{-2/3}$ as Alfv\'{e}n waves interact with weak turbulence. For smaller $M_A$ the range for which weak Alfv\'{e}nic turbulence damps Alfv\'{e}n waves increases. A further increase of the wavelength, i.e. for $\lambda$ from $LM_A$ to $L$, introduces a flat regime of damping, i.e., no dependence on $\lambda$. The damping at this regime is determined by the evolution of turbulence at the injection scale. In its turn, this regime proceeds {\sy untill} $\lambda$ gets of the order of $L$. Finally, {\sy if} $\lambda$ is much larger than $L$, the damping modifies further that it transfers to $\lambda^{-2}$. In that regime the Alfv\'{e}n waves {\sy have so large $\lambda$} that they only feel the distortions that are introduced by turbulence at the outer scale. In comparison, the {\sy FG04} study considered only transAlfv\'{e}nic turbulence and provided only $\lambda^{-1/2}$ scaling for all scales. 

{\sy In terms of the dependence of turbulent damping rate on $M_A$}, it changes from $M_A^2$ for strong turbulence to $M_A^{8/3}$ for weak turbulence. The case of no damping naturally follows as $M_A \rightarrow 0$. As for {\sy FG04} study, it only provided the result for $M_A=1$.

\subsection{Damping by SuperAlfv\'{e}nic turbulence}

As we discussed earlier, if turbulence is superAlfv\'{e}nic, at large scales the effects of magnetic field are marginal and turbulence is {\sy hydrodynamic-like}. However, the turbulent {\sy velocity} decreases with the decreasing scale and at a scale $l_A$ becomes equal to the Alfv\'{e}n velocity. This scale can be considered as the {\sy injection scale} of transAlfv\'{e}nic turbulence. Therefore, the case of Alfv\'{e}n wave damping by superAlfv\'{e}nic turbulence at scales less than $l_{A}$ can be related to the case of damping by transAlfv\'{e}nic
turbulence considered in FG04. Indeed, a simple substitution of $L$ by $l_A$ provides the required rate of magnetic structure evolution on scales less than $l_A$. This gives:
\begin{equation}
\Gamma_{super}\approx \frac{V_A}{l_{A}^{1/2} \lambda^{1/2}}=\frac{V_A M_A^{3/2}}{L^{1/2}\lambda^{1/2}}.
\label{gamma_super}
 \end{equation}

Treating $l_{A}$ as the effective injection scale and using Eq. (\ref{max1}), it is easy to obtain the maximal wavelength up to which
our treatment of the non-linear damping is applicable:
\begin{equation}
\lambda_{max, super}\approx l_{A}=LM_A^{-3}.
\label{max_super}
 \end{equation}
Associating $\lambda$ with $r_L$, we define the corresponding range of
$r_L$
\begin{equation}
\frac{l_{min}^{4/3}}{L^{1/3}} M_A<r_L<LM_A^{-3},
\label{range_super}
 \end{equation}
assuming that the minimal/damping scale of turbulent motions $l_{min}$ is less than $LM_A^{-3}$. 

For Alfv\'{e}n waves with $\lambda$ larger than that given by Eq. (\ref{max_super}) and therefore for $r_L>LM_A^{-3}$, the damping is induced by Kolmogorov-type isotropic hydrodynamic turbulence. The characteristic damping rate in this case is expected to coincide with the  eddy turnover {\sy rate}, i.e.
\begin{equation}
\Gamma_{hydro}\approx \frac{u_{\lambda}}{\lambda}\approx \frac{V_A}{l_A^{1/3} \lambda^{2/3}} \approx \frac{V_A M_A}{L^{1/3} \lambda^{2/3}},
\label{hydro_dam}
 \end{equation}
 where we use Eq. (\ref{u_hydro}). 
 
 Similar to the case of sub-Alfv\'{e}nic turbulence, in superAlfv\'{e}nic case, we observe the change of the rate of Alfv\'{e}n wave damping changing from $\lambda^{-1/2}$ for short wavelengths to $\lambda^{-2/3}$ for $\lambda$ longer than $LM_A^{-3}$. The turbulent damping rate of Alfv\'{e}n waves increases with $M_A$. 
 
\subsection{Other forms of presenting our results}


The scaling of weak turbulence is different from that of strong turbulence that starts at the transition scale $l_{trans}=LM_A^2$ of subAlfv\'{e}nic turbulence. 
However, what is the same in the two regimes of turbulence is the cascading rate. Indeed, the {\sy energy cascades at the same rate} without accumulating at any scale and dissipates only at the small dissipation scale. Therefore, by expressing the dissipation rate of Alfv\'{e}n waves through the cascading rate of turbulence, we will demonstrate a higher degree of universality of the obtained expressions. 

The cascading rate of the weak turbulence is given by Eq. (\ref{eps_weak}) and we can write it as
\begin{equation}
\epsilon_w\approx \frac{V_A^3 M_A^4}{L}.
\label{eps_w2}
 \end{equation}
{This} reflects the decrease of energy dissipation by $M_A^4$ compared to the case of transAlfv\'{e}nic turbulence in FG04. If $r_L<LM_A^4$, the rate of Alven wave damping can be obtained by combining Eq. (\ref{eps_w2}) and Eq. (\ref{gamma1}):
\begin{equation}
\Gamma_{subA, s}\approx \frac{\epsilon_w^{1/2}}{V_A^{1/2} r_L^{1/2}}.
\label{gamma_subA_eps}
 \end{equation}
The peculiar feature of Eq. (\ref{gamma_subA_eps}) is that if one formally substitutes instead of $\epsilon_w$ the cascading rate of strong turbulence, one will get the expression in FG04. This is exactly the universality of expressions that we sought. Nevertheless, this analogy is only formal as the cascading rate for weak turbulence is $M_A^4$ times lower compared to the transAlfv\'{e}nic case. Therefore, the obtained damping rate for subAlfv\'{e}nic turbulence is $M_A^2$ times less than the case of trans-Alfv\'{e}nic turbulence (see also Eq. (\ref{gamma1})). 

For wavelengths in the range $L M_A^4<\lambda<LM_A$ the weak turbulence is responsible for the Alfv\'{e}n wave damping. Thus, expressing $M_A$ from Eq. (\ref{eps_w2}) and substituting it in Eq. (\ref{gamma_weak}) we can get
\begin{equation}
\Gamma_{subA, w}\approx \frac{\epsilon_w^{1/3} L^{1/3}}{V_A r_L^{2/3}}\approx \frac{\epsilon^{1/3}_w M_A^{4/3}}{r_L^{2/3}}.
\label{gamma_weak2}
 \end{equation}
The expression given by Eq. (\ref{gamma_weak2}) demonstrates a slower damping rate in comparison to Eq. (\ref{gamma_subA_eps}). The decrease of damping rate by the factor $M_A^{8/3}$ (see Eq. (\ref{gamma_weak})) is significant. It is important to note that for $M_A\ll 1$ it provides the smooth transition to the regime of marginal Alfv\'{e}n wave damping as the magnetic field perturbations get smaller and smaller. Naturally, this expression is very different from that in FG04, as the latter study did not consider the damping induced by weak Alfv\'{e}nic turbulence. 

For damping of Alfv\'{e}n waves {\sy generated} by CRs with larger $r_L$, i.e. $LM_A<r_L< L$ (see Eq. (\ref{lambda_w})) we obtain:
\begin{equation}
\Gamma_{outer}\approx \frac{\epsilon_w^{1/2}}{L^{1/2} V_A^{1/2}}.
\label{outer}
 \end{equation}

For superAlfv\'{e}nic turbulence at scales less than $l_A$, by expressing $M_A$ from Eq. (\ref{super_alt1}) and substituting it in Eq. (\ref{gamma_super}), we obtain
\begin{equation}
\Gamma_{super}\approx \frac{\epsilon_{super}^{1/2}}{V_A^{1/2}r_L^{1/2}}.
 \end{equation}
Formally, the above expression coincides with the expression for the damping by subAlfv\'{e}nic strong turbulence given by Eq. (\ref{gamma_subA_eps}). Nevertheless, the subAlfv\'{e}nic turbulence demonstrates the significant {\it reduction} of the cascading rate compared to the {\sy transAlfv\'{e}nic turbulence}. On the contrary,  the superAlfv\'{e}nic strong MHD turbulence corresponds to a significant {\it increase} of dissipation rate in comparison with the transAlfv\'{e}nic case.
Thus, for the same injection scale $L$ and the same injection velocity $V_L$,  the damping of Alfv\'{e}n waves depends on the magnetization of media. At a lower magnetization, e.g., for superAlfv\'{e}nic turbulence, the damping of Alfv\'{e}n waves is more efficient than that {\sy at a  higher medium magnetization, i.e. for the subAlfv\'{e}nic case.}

As we discussed earlier, in superAlfv\'{e}nic turbulence, the long Alfv\'{e}n waves with $\lambda$ larger than $l_A=LM_A^{-3}$ interact with hydrodynamic turbulence and the corresponding damping rate is 
\begin{equation}
\Gamma_{hydro}\approx \frac{\epsilon_{hydro}^{1/3}}{r_L^{2/3}},
 \end{equation}
 where the hydrodynamic dissipation rate is $\epsilon_{hydro}\approx V_L^3/L$.

Below we present a few more forms of presenting the damping rates that we obtained above. For instance, 
it could be sometimes useful to rewrite the expressions given by Eq. (\ref{gamma1}) and (\ref{gamma_weak})  in terms of $\lambda_{max, s}$ given by Eq. (\ref{max_s}). We remind the reader that the physical meaning of $\lambda_{max,s}$ is the longest wavelength that still interacts with strong turbulent cascade. Then, 
\begin{equation}
\Gamma_{subA, s}\approx \frac{V_A}{L} \left(\frac{\lambda_{max, s}}{r_L}\right)^{1/2}, ~~~~r_L<\lambda_{max, s},
\label{g_s3}
 \end{equation}
and
\begin{equation}
\Gamma_{subA, w}\approx \frac{V_A}{L} \left(\frac{\lambda_{max, s}}{r_L}\right)^{2/3}, ~~~~r_L>\lambda_{max, s}.
\label{g_w3}
 \end{equation}
It is easy to see that Eq. (\ref{g_s3}) demonstrates that the damping by strong MHD turbulence $\Gamma_{subA, s}$ happens faster than the Alfv\'{e}n crossing rate of the injection scale eddies.  In the case of weak turbulence, Eq. (\ref{g_w3}) demonstrates that $\Gamma_{subA,w}$ is {\sy slower than the above rate}.

\section{Turbulent damping of Alfv\'{e}n waves generated in the global system of reference}
 
{The turbulent damping of Alfv\'{e}n waves {\sy generated} by streaming CRs is an important special case of turbulent damping as the streaming instability induces Alfv\'{e}n waves that are aligned with the local direction of magnetic field. Another case arises if we consider the damping of a flux of Alfv\'{e}n waves {\sy generated} by {\sy an extended source}. The difference between the two cases} is that in the latter setting the waves are {\sy generated} irrespectively to the local direction of magnetic field. Therefore, such Alfv\'{e}n waves should be viewed in the global system of reference related to the mean magnetic field. As a result, our earlier treatment of the Alfv\'{e}n wave damping by MHD turbulence should be modified.  
\subsection{Case of Strong SubAlfv\'{e}nic turbulence}

Consider an Alfv\'{e}n wave {\sy generated} at an angle $\theta\gg \delta B/B$ with respect to the global {\it mean} magnetic field. In this situation it is natural to disregard the dispersion of  angles that arises from magnetic wandering induced by turbulence.\footnote{In the case $\theta\sim \delta B/B$, one should {\sy average} the final expressions over the $\theta$ dispersion that arises from magnetic field wandering.} To distinguish {\sy these two cases} we use $\sin \theta$ instead of $\sin \theta_x$ in Eq. (\ref{alt_lambda}). In this case the perpendicular scale of eddies that the waves {\sy interact with} is given by:
\begin{equation}
x\approx \frac{\lambda}{\sin \theta}.
\label{xglobal}
 \end{equation}
For strong turbulence the rate of the wave damping
is equal to the turnover rate of subAlfv\'{e}nic eddies. Therefore using Eq. (\ref{xglobal}), we find
\begin{equation}
\Gamma_{subA, s, \theta} \approx \frac{V_A M_A^{4/3} \sin^{2/3}\theta}{\lambda^{2/3} L^{1/3}}.
\label{subA_s_global}
 \end{equation}
This provides the non-linear damping rate of an Alfv\'{e}n wave moving at the angle $\theta$ with respect to the mean field.

Using the expression of weak turbulent {\sy cascading rate} $\epsilon_w$ (see Eq. (\ref{eps_weak})), one can write:
\begin{equation}
\Gamma_{subA, s, \theta} \approx \frac{\epsilon^{1/3}_w \sin^{2/3}\theta}{\lambda^{2/3}}.
\label{global_eps}
 \end{equation}
The turbulent damping given by Eq.(\ref{global_eps}) is applicable to 
\begin{equation}
l_{min}\sin\theta < \lambda < LM_A^2 \sin\theta,
\label{range_new}
 \end{equation}
where $l_{min}$ is the perpendicular damping scale, and $LM_A^2=l_{trans}$ is the transition scale from strong to weak MHD turbulence.

Naturally,  the adopted approximation $\theta \gg \delta B/B$ fails if the wave is launched parallel to the mean magnetic field. The directions of 
the local magnetic field deviates from the mean field and this makes the {\sy actual $\theta_0$} different from zero. In the global system of reference the
{\sy dispersion} is dominated by the magnetic field variations {\sy presented at the injection scale} (see Cho et al. 2002). Therefore
\begin{equation}
\theta_0\approx \frac{B_L}{B}\approx M_A.
\label{theta0}
 \end{equation}
Substituting this into Eq. (\ref{subA_s_global}) we get 
\begin{equation}
\Gamma_{subA, s, 0}\approx \frac{\epsilon^{1/3}_w M_A^{2/3}}{\lambda^{2/3}}.
\label{gamma_s_0}
 \end{equation}
The above expression is different from {\sy Eqs. (\ref{gamma1}) and (\ref{gamma_subA_eps})}. The
difference stems from the {\sy different properties of Alfv\'{e}n waves} generated in the local system versus the global system of
reference. 
The damping rate in Eq. (\ref{gamma_s_0}) is applicable to the range of wavelength
\begin{equation}
l_{min} M_A < \lambda < LM_A^3,
 \end{equation}
the {\sy latter result} trivially follows from Eqs.(\ref{range_new}) and (\ref{theta0}).  
\subsection{The case of Weak SubAlfv\'{e}nic turbulence}

For weak subAlfv\'{e}nic turbulence, in the case $\theta\gg \delta B/B$, one should use Eq. (\ref{xglobal}) to relate $\lambda$ to
the scale of perpendicular motions that the wave strongly non-linearly interacts with. To obtain the damping rate, Eq. (\ref{gg}) should be used:
\begin{equation}
\Gamma_{weak, global, \theta} \approx \frac{V_A \sin\theta M_A^2}{\lambda}\approx \frac{\epsilon^{1/2} L^{1/2} \sin\theta}{V_A^{3/2} \lambda},
\label{global_weak}
 \end{equation}
where we use the weak cascading rate $\epsilon_w$.
The range of wavelength for this type of damping is 
\begin{equation}
LM_A^2 \sin\theta< \lambda < LM_A \sin\theta.
\label{range_new2}
 \end{equation}
The last inequality is obtained by substituting the maximal perpendicular eddy scale $LM_A$ for $x$ in 
Eq.(\ref{xglobal}).

In the case of Alfv\'{e}n wave propagation along the mean magnetic field, one should use Eq. (\ref{theta0}) to get
\begin{equation}
\Gamma_{weak, global, 0}\approx \frac{V_A M_A^3}{\lambda}\approx \frac{V_A \epsilon^{3/4} L^{3/4}}{\lambda V_A^{5/4}}.
\label{global_weak_0}
 \end{equation}
Using Eqs.(\ref{theta0}) and (\ref{range_new2}), we find the range of wavelength that is subject to the turbulent damping:  
\begin{equation}
LM_A^3 < \lambda <LM_A^2.
 \end{equation}

 \subsection{Other cases}
 
 After illustrating the difference of non-linear damping for waves generated {\sy in the local reference system of magnetic field and in the global reference system of the mean field,}
 we can provide results for other cases. More detailed discussion was presented in Lazarian (2016). For instance, for superAlfv\'{e}nic turbulence, there is
\begin{equation}
 \Gamma_{super, global, \theta}\approx \frac{V_A M_A \sin^{2/3}\theta}{\lambda^{2/3} L^{1/3}},
 \label{global_super}
 \end{equation}
 where in superAlfv\'{e}nic turbulence angle $\theta$ varies from one turbulent eddy of size $l_A$ to another. As a result, the corresponding averaging over such changing directions should be performed. For the random distribution of the relevant directions, the corresponding geometric factor is $\langle \sin^{2/3} \theta \rangle =3/5$.

 On scales larger than $l_A$, MHD turbulence is marginally affected by magnetic fields. As a result, no difference between local and global frames is present in terms of Alfv\'{e}n wave damping. This difference also disappears for the damping by turbulent fluctuations at the injection scale.
 
 \subsection{Summary of main results in Sections 3 and 4 on turbulent damping}
 
 Some of our results for non-linear turbulent damping of Alfv\'{e}n waves {\sy in different turbulence regimes are summarized} in Table 1.  
 We show results relevant both to the damping of waves in the local system of reference, e.g. corresponding to the waves generated by streaming instability (fifth column in Table 1 with the name ``Instability damping rate"), and the damping of waves {\sy generated} by external sources {\it {\sy parallel to} the mean magnetic field} (sixth column in Table 1 with the name ``Wave damping rate). The table illustrates that the rate of damping and the ranges of wavelengths for which  damping is applicable are very different for the two situations. {\sy At the first glance, this seems strange.} However, the difference stems from the fact that in the case of streaming instability the waves are aligned with the local magnetic field, while the waves generated by an {\sy extended} source are sent {\sy parallel to} the mean magnetic field. 
 
We did not cover in Table 1 {\sy the general case of Alfv\'{e}n waves generated at an arbitrary angle relative to the mean magnetic field}, as well as damping of Alfv\'{e}n waves by outer-scale turbulence.  
{It is also necessary to stress again} the important role of weak turbulence for the suppression of streaming instability at low $M_A$. While the weak turbulence is frequently disregarded due to its  limited {\sy inertial range $[LM_A^2, L]$}, it can affect CR streaming for $r_L$ in the range $[LM_A^4, LM_A]$, which can be extensive for sufficiently small $M_A$.

\begin{figure*}
\centering
\includegraphics[width=18cm]{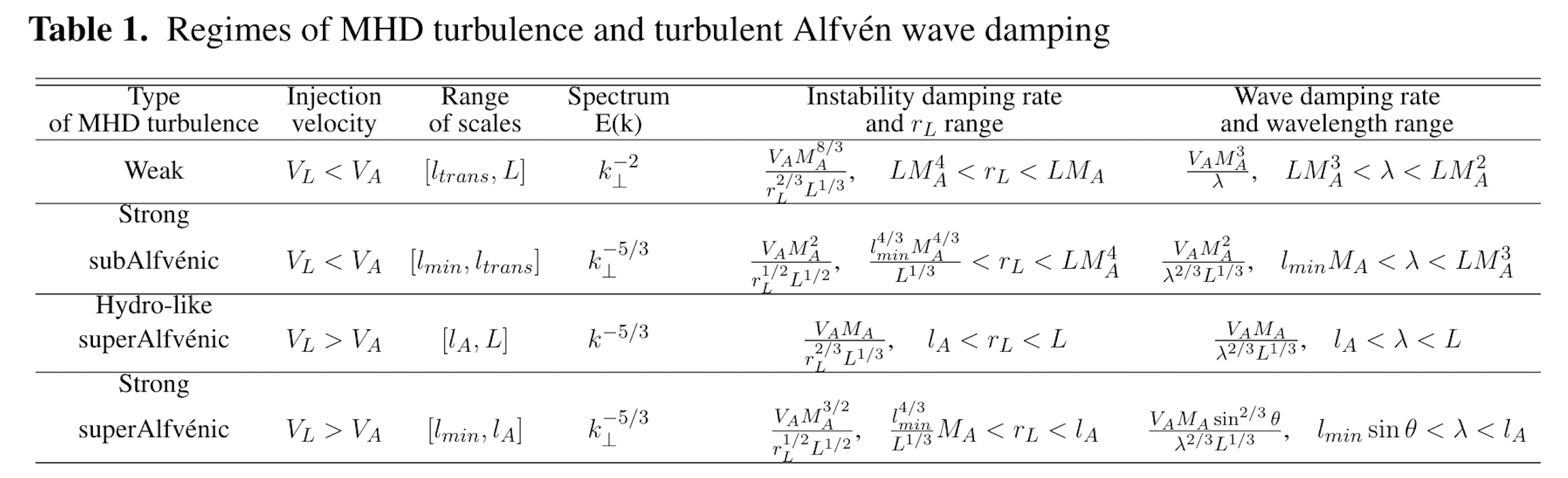}
\end{figure*}

\section{Ion-neutral collisional damping of streaming instability}

In the  presence of partial ionization, an additional effect of damping by ion-neutral collisions becomes important. This effect was discussed originally by Kulsrud \& Pearce (1965) for Alfv\'{e}n waves. The damping of turbulent motions in partially ionized gas was recently summarized in Xu \& Lazarian (2017). 

In the presence of neutrals, a slippage between them and ions induces the dissipation. In a mostly neutral medium, at wave frequencies $\omega = V_A k_\|$ less than the neutral-ion collisional frequency $\nu_{ni}$, both species move together and the dissipation is minimal. As the wave frequency increases, not all neutrals get the chance to collide with ions and the relative motions of ions and neutrals induce significant dissipation.
For strongly coupled ions and neutrals, 
the ion-neutral collisional (IN) damping rate is
(Kulsrud \& Pearce 1969)
\begin{equation}\label{eq: indamgen}
   \Gamma_{IN} =  \frac{\xi_n V_A^2 k_\|^2}{2\nu_{ni}},
\end{equation}
where  
$\xi_n = \rho_n /\rho$, 
and $\rho_n$ and $\rho$ are the neutral and total mass densities. 
For weakly coupled ions and neutrals with $\omega = V_{Ai} k_\| > \nu_{in}$, 
where $V_{Ai}$ is the Alfv\'{e}n speed in ions and $\nu_{in}$ is the ion-neutral collisional frequency,  
there is 
\begin{equation}\label{eq: indrwc}
   \Gamma_{IN} = \frac{\nu_{in}}{2}.
\end{equation}
We note that both turbulent and wave motions are subject to the IN damping. 
Strong Alfv\'{e}nic turbulence injected in the strong coupling regime cannot cascade into the weak coupling regime due to the severe damping effect
(Xu et al. 2015, 2016).


IN damping 
is sensitive to the ionization fraction and becomes weak at a high ionization fraction. 
For strongly coupled ions and neutrals with $V_A k_\| < \nu_{in}$, 
$\Gamma_{IN}$ is still given by Eq. \eqref{eq: indamgen}.
For decoupled ions with $V_{Ai} k_\| > \nu_{in}$, there is 
(Xu et al. 2016)
\begin{equation}
    \Gamma_{IN} = \frac{\nu_{ni} \chi   V_{Ai}^2 k_\|^2}{2 \big[(1+\chi)^2 \nu_{ni}^2 +  V_{Ai}^2 k_\|^2\big]}, 
\end{equation}
where $\chi = \rho_n /\rho_i$ and $\rho_i$ is the ion mass density. Furthermore, when 
neutrals are also decoupled from ions with 
$V_{Ai}k_\| > \nu_{ni}$, the above expression is reduced to Eq. \eqref{eq: indrwc}. 
Because of the weak damping effect, 
Alfv\'{e}nic cascade in a highly ionized medium is not dissipated by IN damping
(Xu \& Lazarian 2022).

Naturally, to understand whether turbulent damping or IN damping is more important for damping the streaming instability, $\Gamma_{IN}$ should be compared with the turbulent damping rate $\Gamma$ that we provided earlier. This comparison has been recently carried out in detail by 
Xu \& Lazarian (2022).
Here we selectively review some of their results. 

In a weakly ionized interstellar medium, e.g., molecular clouds, CR-driven Alfv\'{e}n waves are likely in the weak coupling regime with
\begin{equation}\label{eq: cuofrwcrl}
   \frac{V_{Ai}}{r_L \nu_{in}} \approx  2\times10^3 \Big(\frac{B_0}{10~\mu\text{G}}\Big)^2 
   \Big(\frac{n_H}{100~\text{cm}^{-3}}\Big)^{-\frac{3}{2}} \Big(\frac{n_e/n_H}{10^{-4}}\Big)^{-\frac{1}{2}} 
   \Big(\frac{E_{CR}}{1~\text{GeV}}\Big)^{-1} \gg 1,
\end{equation}
where $B_0$ is the mean magnetic field strength, $n_e$ and $n_H$ are number densities of electrons and atomic hydrogen, and $E_{CR}$ is the CR energy. 
As already mentioned above, 
strong Alfv\'{e}nic turbulence injected at a large scale in the strong coupling regime is severely damped and its cascade cannot persist in the weak coupling regime. Therefore, there is 
\begin{equation}
   \Gamma < \Gamma_{IN} = \frac{\nu_{in}}{2}.
\end{equation}
So the damping of streaming instability in a weakly ionized medium is dominated by IN damping. 

In a highly ionized interstellar medium, e.g., the warm ionized medium,
CR-generated Alfv\'{e}n waves are still in the weak coupling regime and have 
\begin{equation}
    \frac{V_{Ai}}{r_L \nu_{ni}} 
    = 7.6\times10^3 
    \Big(\frac{B_0}{1~\mu\text{G}}\Big)^2 \Big(\frac{n_i}{0.1~\text{cm}^{-3}}\Big)^{-\frac{3}{2}} \Big(\frac{E_{CR}}{1~\text{GeV}}\Big)^{-1} 
    \gg1.
\end{equation}
To have the turbulent damping dominate over IN damping, 
there should be 
\begin{equation}
      \frac{\Gamma}{\Gamma_{IN}} = \frac{\Gamma}{\frac{\nu_{in}}{2}} >1,
\end{equation}
which can be rewritten as 
\begin{equation}\label{eq: contuinsup}
\begin{aligned}
   M_A &> \Big(\frac{\nu_{in}}{2} V_{Ai}^{-1} L^\frac{1}{2} r_L^\frac{1}{2}\Big)^\frac{2}{3} \\
          &= 0.2 \Big(\frac{B_0}{1~\mu\text{G}}\Big)^{-1} \Big(\frac{n_i}{0.1~\text{cm}^{-3}}\Big)^{\frac{1}{3}} 
          \Big(\frac{n_n}{0.01~\text{cm}^{-3}}\Big)^{\frac{2}{3}} 
          \Big(\frac{L}{100~\text{pc}}\Big)^\frac{1}{3}
          \Big(\frac{E_{CR}}{1~\text{GeV}}\Big)^{\frac{1}{3}} 
\end{aligned}
\end{equation}
for superAlfv\'{e}nic turbulence, 
where $n_i$ and $n_n$ are the number densities of ions and neutrals, 
and
\begin{equation}
\begin{aligned}
   M_A &> \Big(\frac{\nu_{in}}{2} V_{Ai}^{-1} L^\frac{1}{2} r_L^\frac{1}{2}\Big)^\frac{1}{2} \\
            &= 0.3 \Big(\frac{B_0}{1~\mu\text{G}}\Big)^{-\frac{3}{4}} \Big(\frac{n_i}{0.1~\text{cm}^{-3}}\Big)^{\frac{1}{4}} 
          \Big(\frac{n_n}{0.01~\text{cm}^{-3}}\Big)^{\frac{1}{2}} 
          \Big(\frac{L}{100~\text{pc}}\Big)^\frac{1}{4}
          \Big(\frac{E_{CR}}{1~\text{GeV}}\Big)^{\frac{1}{4}} 
\end{aligned}
\end{equation}
for subAlfv\'{e}nic turbulence.
We see that the condition in Eq. \eqref{eq: contuinsup} is naturally satisfied for superAlfv\'{e}nic turbulence. 
In a highly ionized medium, as the IN damping is weak, streaming instability is predominantly damped by the turbulent damping.

\begin{figure}
\centering
\includegraphics[width=16cm]{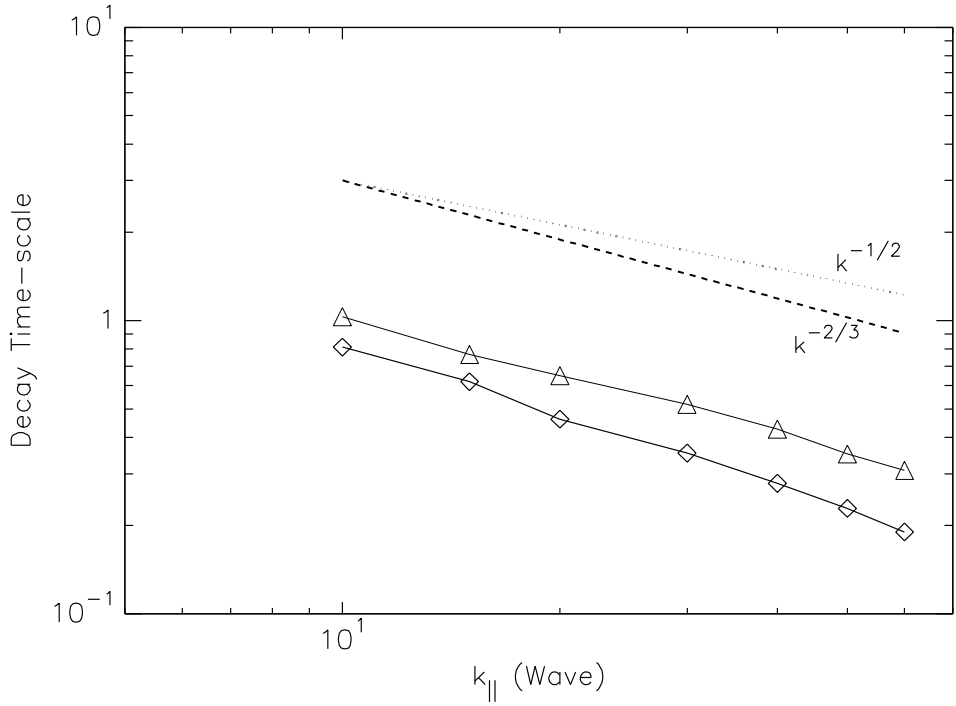}
\caption{The damping time-scale $\Gamma^{-1}$ of Alfv\'{e}n waves that are injected at {\sy $k_\|=10$ in 3D MHD turbulence}, where the parallel direction is chosen with respect to the {\it mean} magnetic field. In one approach the Alfv\'{e}n wave energy $E_{w}$ decays in the turbulent medium over the time scale 
$\tau_1=\ln(E(t_1)/E(t_2))/(t_2-t_1)$. The values of $\tau_1$ are given by triangular symbols. 
In the other approach the wave energy is {\sy continuously injected at $k_\|=10$} until it reaches a saturation level $E_w$. The {\sy corresponding damping time scale} is given by $\tau_2=E_w/\epsilon_{driving}$, where $\epsilon_{driving}$ is the wave energy injection rate. $\tau_2$ is denoted by diamond symbols. 
{\sy The two measurements are both consistent with $k^{-2/3}$ scaling.} From Cho \& Lazarian 2022.}
\label{damping}
\end{figure}

\section{Numerical testing of turbulent damping of Alfv\'{e}n waves}

Numerical testing of Lazarian (2016) is essential in {\sy a variety of regimes}. 
{\sy By using 3D MHD turbulence simulations (Cho et al. 2002)},
the results of numerical testing {\sy on turbulent damping of externally driven Alfv\'{e}n waves} are presented in Figure \ref{damping}. The observed scaling is consistent with Lazarian (2016) predictions, but inconsistent with FG04 prediction. 

The reason for this difference arises from {\sy the global reference frame adopted in} the numerical experiment. Launching of Alfv\'{e}n waves with respect to the local direction of magnetic field is complicated in turbulent fluid. Therefore, the testing presented in Figure \ref{damping} was {\sy carried out} with Alfv\'{e}n waves launched with respect to the mean magnetic field. This is the setting corresponding to turbulent 
damping of Alfv\'{e}n waves generated in the global system of reference that we considered in {\sy \S 4}. As a result, the numerical simulations confirmed the scaling of {\sy inverse of damping rate 
$\Gamma^{-1}$, i.e., damping time scale}, 
{\sy which is measured at different $\lambda$ as $\lambda^{2/3}\sim k_\|^{-2/3}$}. This result is different from the  prediction of $\Gamma^{-1}\sim k_\|^{-1/2}$ {\sy of streaming instability} in FG04 for transAlfv\'{e}nic turbulence and in
Lazarian (2016) for the strong Alfv\'{e}nic turbulence part of the cascade for a wide range of $M_A$.  {\sy Numerical testing on turbulent damping of streaming instability in the local reference frame requires a more complicated setup and has not been performed so far. }
 
 \section{Astrophysical Implications}

 \subsection{Propagation of CRs}

 For decades the study on CR propagation was performed within a simple model, the so-called ``leaky box model"  (see Longair 2011). In this model Galactic CRs propagate freely within the partially ionized disk of the Galaxy. The Alfv\'{e}n waves experience damping in the partially ionized gas (Kulsrud \& Pearce 1969, Lithwick \& Goldreich 2001, Xu et al. 2016, 2017) and thus the streaming instability is suppressed. 
On the contrary, in fully ionized plasmas of the Galactic halo, the damping of Alfv\'{e}n waves is significantly reduced and the streaming instability is present. Therefore, in this classical simplistic picture that ignores turbulence, Galactic CRs stream freely through the Galactic disk and  are scattered backwards in the Galactic halo.

This classical ``leaky box model" is {\sy problematic}, as it is well known now that the Galactic disk is not fully filled with partially ionized gas. 
In fact, a significant fraction of the Galactic disk material is {\sy warm} ionized gas (McKee \& Ostriker 1977, Draine 2011). Therefore, CRs cannot {\sy zoom through} the Galactic disk due to the streaming instability. 


FG04 quantified the idea of turbulent damping of streaming instability mentioned in Yan \& Lazarian (2002) and came to a paradoxical conclusion by applying their theory to the propagation of CRs in the Galaxy.
{\sy By assuming homogeneous transAlfv\'{e}nic turbulence in the Galaxy, they found significant turbulent damping of streaming instability and thus poor confinement of CRs.}
This would entail problems with explaining e.g., the observed isotropy of CRs and their residence time in the Galaxy. 




In Lazarian (2016) the gist of the ``leaky box model" was preserved, but instead of damping by ion-neutral collisional friction, the study appealed to the turbulent damping of streaming instability in the Galactic disk and proposed a ``turbulent leaky box model". 
{\sy Different from FG04, by considering inhomogeneous turbulence properties in the Galaxy and the strong $M_A$ dependence of turbulent damping, they found that the damping by weak subAlfv\'{e}nic turbulence is marginal in the Galactic halo and thus CRs, even at high energies, can still be confined by streaming instability.  }

\begin{figure}[ht]
\centering
   \includegraphics[width=9cm]{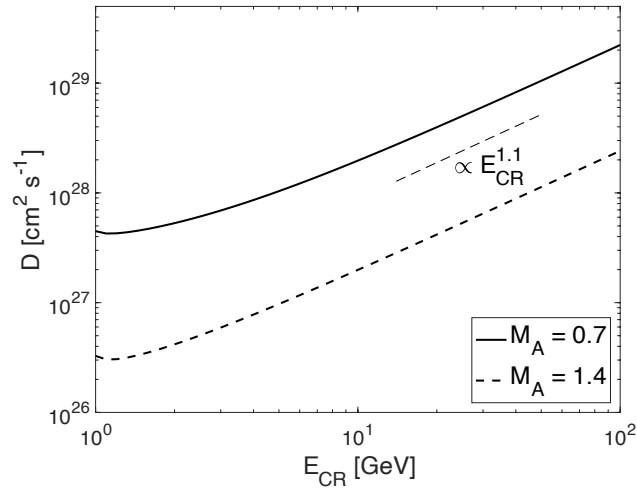}
\caption{Diffusion coefficient vs. $E_{CR}$ of streaming CRs
in super and subAlfv\'{e}nic turbulence in the 
WIM.
From Xu \& Lazarian (2022).}
\label{fig: dffwim}
\end{figure}

{\sy In a recent study by Xu \& Lazarian (2022), they identified the important role of turbulent damping of streaming instability in the warm ionized medium (WIM).
Fig. \ref{fig: dffwim} shows the diffusion coefficient $D$ of streaming CRs. The $M_A$ dependence comes from both turbulent damping of streaming instability and wandering of turbulent magnetic field lines.
In particular, 
the smaller $D$ in superAlfv\'{e}nic turbulence is caused by the tangling of turbulent magnetic fields, which results in an effective mean free path $l_A$ of the CRs streaming along turbulent magnetic fields
(Brunetti \& Lazarian 2007).}

{\sy The $M_A$-dependent diffusion of CRs is important for a realistic modeling of inhomogeneous CR diffusion in the Galaxy
(Xu 2021).}
The actual values of $M_A$ in the Galaxy can be measured from observations using a newly developed gradient technique (Lazarian et al. 2018, see also Xu \& Hu 2021) or with more traditional magnetic field and turbulent velocity measurements.

\subsection{Launching of winds and heating}

While the damping of Alfv\'{e}n waves by turbulence is an accepted process in the field of CR research,  we would like to point out that the turbulent damping of Alfv\'{e}n waves can be responsible for many fundamental astrophysical processes. For instance, different processes of damping were discussed for heating of stellar corona by Alfv\'{e}n waves, as well as for launching of stellar winds (see Suzuki \& Inutsuka 2005, Verdini et al. 2005, Evans et al. 2009, Vidotto \& Jatenco-Pereira 2010, Verdini et al. 2010, Suzuki 2015). It is clear that the turbulent damping of Alfv\'{e}n waves can be very important in these settings. More recently,  launching galactic winds by {\sy turbulent damping of the Alfv\'{e}n waves} generated by galactic activity was considered in Suzuki \& Lazarian (2017). 
Accounting for the dependence of turbulent damping on $M_A$ is important for the quantitative modeling of the process. A similar process is important for launching winds from other types of active disk systems, e.g. circumstellar disks. 

Apart from launching galactic winds by turbulent damping of Alfv\'{e}n waves generated by the galaxy, the turbulent damping of {\sy streaming instability} also plays a very important role in coupling CRs and magnetized galactic matter. 
The pressure of CRs in galactic settings is significant and it can modify interstellar dynamics. Galactic winds driven by CRs present an important example of this modification. 

In general, the importance of galactic winds is easy to understand. For galaxies of the Milky Way luminosity, about 20 percent of baryons are accounted for when matching the observed luminosity to the halo mass function. Observing absorption lines in spectra of background quasars testifies  for the efficient expulsion of galactic baryons from the galaxies. In fact there is evidence that galaxies with significant star formation can drive mass outflows up to 10 times the rate of star formation (Brand-Hawthorn et al. 2007). 

Numerical simulations have demonstrated that CRs indeed influence the generation of global outflows and the local structure of the interstellar medium (ISM) (see Ruszkowski et al. 2017). The exact properties of the simulated outflows depend sensitively on how CR transport is modeled. Recent simulations by Holguin et al. (2019) employed Lazarian (2016) model of turbulent damping and obtained the results that differ significantly from the earlier modeling in e.g., Ruszkowski et al. (2017). The difference stemmed from the fact that the earlier calculations employed the model by FG04, which is only {\sy applicable to} transAlfv\'{e}nic turbulence, i.e. $M_A=1$. {\sy However}, the actual $M_A$ of gas can  vary significantly in simulations. 

\begin{figure*}
\centering
\includegraphics[width=17cm]{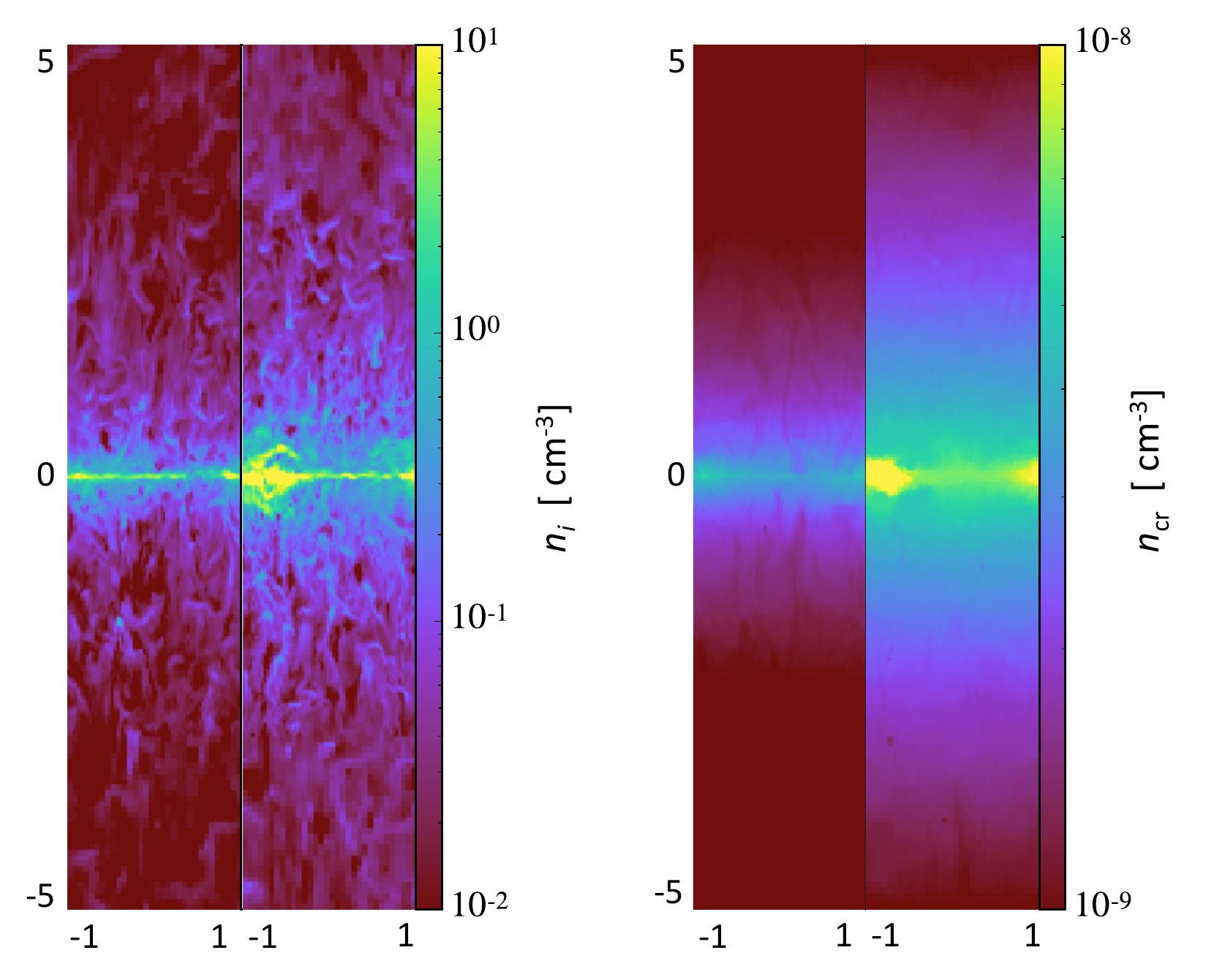}
\caption{Simulations of the galactic ISM evolution in the presence of star formation and CR driven outflows. The figure shows {\sy the gas ($n_i$) and CR ($n_{cr}$) density slices}  $\pm 5 $ kpc along  $z$ direction perpendicular to the midplane obtained in two simulations over time  200 Myr. The CR streaming is affected by turbulent damping of streaming instability with {\sy the turbulent velocity} $\sigma  = 10$ km/s. The results obtained in the absence of turbulent damping on the left side of each pair of plots 
{\sy are clearly different from those with turbulent damping on the right side.} 
The distribution of {\sy both gas and CRs} is more extended in the presence of turbulent damping.  
From Hoguin et al. (2019). }
\label{density1}
\end{figure*}

The results of the numerical simulations in Holguin  et al. (2019) are presented in Figure \ref{density1}. Some of the implications 
include, fist of all,
{\sy when turbulent damping of CR streaming instability is included, 
there is an increase of star formation rate, and the increase is more significant at a higher level of turbulence. 
The reason is that} the turbulent damping increases the average CR streaming speed. This allows CRs to leave the dense mid-plane, reducing the pressure support from CRs to the gas. As a result, the gas in the disk collapses and stars form more {\sy efficiently}. 
{\sy Furthermore}, the higher efficiency of star formation results in more CRs produced in the mid-plane. {\sy The increased streaming speed of CRs leads to a more extended CR distribution away from the mid-plane}.
It is also important that the escape of CRs from the dense regions allows them to interact with lower-density gas. This  widens the gas distribution in height and accelerates the gas to form CR-driven galactic winds.

In addition, the theory of Alfv\'{e}n wave damping by turbulence suggests that Alfv\'{e}n waves can propagate across longer distances in highly magnetized regions of solar atmosphere (small $M_A$) compared to the regions with higher $M_A$. 
This prediction can be observationally tested. 
This effect should be accounted in both modelling of solar wind launching and modelling of plasma heating. For instance, it is likely that the turbulent damping can be important in order to explain the observed ``unexpected" damping of Alfv\'{e}n waves in the regions above the Sun's polar coronal holes (Hahn et al. 2012). 





\section{Summary}

Alfv\'{e}n waves are damped in turbulent media and the damping depends on the Alfv\'{e}n Mach number $M_A$ of the turbulence. 
At the same wavelength, the wave damping depends on whether the waves are generated 
in the local reference system of magnetic eddies by the CR streaming or they are injected at an angle relative to the {\sy large-scale} mean magnetic field from an {\sy extended} astrophysical source. The latter is, e.g., the case of the Alfv\'{e}n waves arising from magnetic reconnection, {\sy or oscillations} in accretion disks and stellar atmospheres. 
The difference in their damping rates arises from the difference between the local and global systems of reference where the Alfv\'{e}n waves are {\sy generated}. 

The dependence of damping rate on the wavelength $\lambda$ of the Alfv\'{e}n waves in the local system of reference is $\lambda^{-1/2}$, as opposed to a stronger dependence $\lambda^{-2/3}$ for the waves
in the global reference system.

The turbulent damping also depends on whether Alfv\'{e}n waves interact with weak or strong Alfv\'{e}nic turbulence. For $M_A<1$, the turbulence from the injection scale $L$ to the scale $LM_A^2$ is weak and is strong at smaller scales. 
{\sy Weak turbulence can play an important role
in turbulent damping of streaming instability driven by high-energy CRs at a small $M_A$. }


{\sy In a partially ionized gas, the turbulent damping still dominates 
the damping of streaming instability
when the ionization fraction is sufficiently high, 
e.g., in the warm ionized medium
(Xu \& Lazarian 2022).
In star burst galaxies, the ionization fraction is low and the ion-neutral collisional damping can be more important
(e.g., Krumholz et al. 2020).}


{\sy The turbulent damping of streaming instability has important implications on propagation of CRs in the Galaxy, star formation, coupling between CRs and magnetized gas and thus driving galactic winds. 
In addition, }
the turbulent damping of Alfv\'{e}n waves results in heating of the medium and {\sy transfer of the momentum from Alfv\'{e}nic flux to the medium}. {\sy The latter is also important for launching winds. }

\acknowledgments
The research is supported by NASA TCAN 144AAG1967. 
S.X. acknowledges the support for 
this work provided by NASA through the NASA Hubble Fellowship grant \# HST-HF2-51473.001-A awarded by the Space Telescope Science Institute, which is operated by the Association of Universities for Research in Astronomy, Incorporated, under NASA contract NAS5-26555. 
A.L. thanks Jungyeon Cho for discussing the paper.

\bibliographystyle{aasjournal}




\end{document}